\documentclass[12pt,twoside]{article}
\usepackage{amsmath} 

\newcommand{\ts}{\thinspace} 
\newcommand{\lips}{\ts.\ts.\ts.\ts} 
\newcommand{\plips}{.\ts.\ts.\ts} 

\begin{document}

\title{\phantom{$^*$}Fermi and the Elucidation of Matter\thanks{Opening talk
at the celebration of Fermi's 100th birthday, Chicago, September 29, 2001.  To be
published in \emph{Fermi Remembered}, ed. J. Cronin (University of Chicago
Press).}}
\author{Frank Wilczek\\[0.75ex]
\small\it Center for Theoretical Physics\\ 
\small\it Massachusetts Institute of Technology\\ 
\small\it Cambridge, MA 02139-4307}
\date{\small MIT-CTP\kern-.06em -3227}

\maketitle

\thispagestyle{empty}

\markboth{Frank Wilczek}{Fermi and the Elucidation of Matter}

\begin{abstract}\noindent 
Fermi helped establish a new framework for
understanding matter, based on quantum theory.  This framework refines
and improves  traditional atomism in two crucial respects.   First, the
elementary constituents of matter belong to a very small number of classes, and
all objects of a given class (e.g., all electrons) are rigorously identical, indeed
indistinguishable.  This profound identity is demonstrated empirically through
the phenomena of quantum statistics, and is explained by the principles of free
quantum field theory.  Second, objects of one class can mutate into objects of
other classes.  Such mutability can be understood as manifesting
interacting quantum field theory.  Fermi contributed to establishing theoretical
foundations for the new viewpoint,  through his work on quantum statistics and
quantum field theory, and to its fruitful application and empirical validation,
through his work on beta decay, nuclear transmutation, and primeval strong
interaction theory.  
\end{abstract}
\bigskip

\noindent
I feel privileged to return to my alma mater to pay tribute to the memory of
Enrico Fermi on the occasion of the 100th anniversary of his birth.  Fermi has
always been one of my heroes, as well as my scientific ``great-grandfather'' (in
the line Fermi$\rightarrow$Chew$\rightarrow$Gross$\rightarrow$FW).   It was
also a joy and an inspiration, in preparing, to browse through his papers.

I was asked to speak for one-half hour on ``Fermi's Contribution to Modern
Physics''.  This task requires, of course, severe selection.  Later speakers will
discuss Fermi's remarkable achievements as teacher and scientific statesman, so
clearly my job is to focus on his direct contributions to the scientific literature.  
That still leaves far too much material, if I'm to do anything but catalogue. 
Cataloguing would be silly, as well as tedious, since Fermi's collected works, with
important commentaries from his associates, are readily available~\cite{works,
segre}.   What I decided to do, instead, was to try  identifying and following 
some unifying thread that could tie together Fermi's most important work.   
Though Fermi's papers are extraordinarily varied and always focused on specific
problems, such a thread was not difficult to discern.   Fermi was a prolific
contributor to what I feel is the most important achievement of
twentieth-century physics: an accurate and, for practical purposes, completely
adequate theory of matter, based on extremely nonintuitive concepts, but
codified in precise and useable equations.

\section{Atomism Transformed}

Since the days of Galileo and Newton, the goal of physics --
rarely articulated, but tacit in  its practice -- was to derive dynamical equations
so that, given the configuration of a material system at one time, its configuration
at other times could be predicted.   The
description of the Solar System, based on Newtonian celestial mechanics,
realzed this goal.  This description excellently accounts for Kepler's laws of
planetary motion, the tides, the precession of the equinoxes, and much else, but
it gives no  a~priori predictions for such things as the number of planets
and their moons, their relative sizes, or the dimensions of their orbits.   Indeed,
we now know that other stars support quite different kinds of planetary systems.
Similarly, the great eighteenth- and nineteenth-century discoveries in electricity,
magnetism, and optics, synthesized in Maxwell's dynamical equations of
electromagnetism, provided a rich description of the behavior of given
distributions of charges, currents, and electric and magnetic fields, but no
explanation of why there should be specific reproducible forms of matter.   

More specifically, nothing in classical physics explains the existence of
elementary building blocks with definite sizes and properties.   
Yet one of the most basic facts about the physical world is that matter is built up
from a few fundamental building blocks (e.g., electrons, quarks, photons,
gluons), each occurring in vast numbers of identical copies.  Were this not true
there could be no lawful chemistry, because every atom would have its own
quirky properties.   In Nature, by contrast, we find accurate uniformity of
properties, even  across cosmic scales.  The patterns of spectral lines
emitted by atoms in the atmospheres of stars in distant galaxies match those we
observe in terrestrial laboratories.  

Qualitative and semiquantitative evidence for some form of atomism has been
recognized for centuries~\cite{atoms}.  Lucretius gave poetic expression to
ancient atomism and Newton endorsed it in his famous Query 31, beginning
\begin{quote} It seems probable to me, that God in the beginning formed matter
in solid, massy, hard, impenetrable, moveable particles, of such sizes and figures,
of such sizes and figures, and with such other properties, and in such
proportions to space, as most conduced to the ends for which He formed them;
and that these primitive particles being solids, are incomparably harder than any
porous bodies compounded of them, even so very hard, as never to wear or
break in pieces; no ordinary power being able to divide what God Himself made
one in the first creation.
\end{quote} In the nineteenth century Dalton, by identifying the law of definite
proportions, made atomism the basis of scientific chemistry.  Clausius, Maxwell,
and Boltzmann used it to construct successful quantitative theories of the
behavior of gases.    But in these developments the properties of atoms
themselves were not derived, but assumed, and their theory was not much
advanced beyond Newton's formulation.  In particular, two fundamental
questions went begging.  

\paragraph{Question  1:} Why is matter built from vast numbers of particles
that fall into a small number of classes? And why do all particles of a given class
rigorously display the same properties?
 \bigbreak

The indistinguishability of particles is so familiar and so fundamental  to all of
modern physical science  that we could easily take it  for granted.   Yet it is by
no means obvious.   For example, it directly contradicts one of the pillars of
Leibniz's metaphysics, his ``principle of the identity of indiscernibles,''
according to which two objects cannot differ solely in number, but will always
exhibit some distinguishing features.   Maxwell thought the similarity of different
molecules so remarkable that he devoted the last part of his {\it Encyclopedia
Britannica\/} entry on Atoms -- well over a thousand words -- to discussing it.  He
concluded 
\begin{quote} the formation of a molecule is therefore an event not belonging to
that order of nature in which we live\lips it must be referred to the epoch, not of
the formation of the earth or the solar system\lips but of the establishment of the
existing order of nature\plips. \end{quote}

\paragraph{Question  2:}   Why are there different classes of particles?   And why
do they exist in the proportions they do?
\bigbreak

As we have just seen, both Newton and Maxwell considered this question, but
thought that its answer laid beyond the horizon of physics.       

By the end of the twentieth century, physics had made decisive progress on these
questions.    From a science of ``how'', it had expanded into a science of ``what''
that supported a much deeper appreciation of ``why''.  A radically new model of
matter had been constructed.  The elementary building-blocks had been
inventoried, the equations for their behavior had been precisely defined.   The
building-blocks of our transformed atomism are, paradoxically, both far more
reproducible and far more fluid in their properties than classically
conceived atoms.    Fermi was a chief architect of this construction, contributing
to the design at many levels, as I shall now discuss. 

\section{The Identical and the Indistinguishable}

From the perspective of classical physics the indistinguishability of electrons (or
other elementary building blocks) is both inessential and surprising.  If electrons
were nearly but not quite precisely identical,  for example, if their masses
varied over a range of a few parts per billion, then according to the laws of
classical physics different specimens  would behave in nearly but not quite the
same way.   And since the possible behavior is continuously graded, we could not
preclude the possibility that future observations, attaining greater accuracy than
is available today, might discover small differences among electrons.    Indeed, it
would seem reasonable to expect that differences would arise, since over a long
lifetime each electron might wear down, or get bent, according to its
individual history.

The first evidence that the similarity of like particles is quite precise, and goes
deeper than mere resemblance, emerged from a simple but profound reflection
by Josiah Willard Gibbs in his work on the foundations of statistical mechanics. 
It is known as ``Gibbs' paradox'', and goes as follows.    Suppose that we have a
box separated into two equal compartments A and B, both filled with equal
densities of hydrogen gas at the same temperature.  Suppose further that there is
a shutter separating the compartments, and consider what happens if we open
the shutter and allow the gas to settle into equilibrium.   The molecules originally
confined to A (or B) might then be located anywhere in A\ts+\ts B.    Thus, since
there appear to be many more distinct possibilities for distributing the
molecules, it would seem that the entropy of the gas, which measures the
number of possible microstates, will increase.  On the other hand, one might
have the contrary intuition, based on everyday experience, that the properties of
gases in equilibrium are fully characterized by their volume, temperature, and
density.  If that intuition is correct, then in our
thought experiment the act of opening the shutter makes no change in the state of the gas, and so of course it
generates no entropy.   In fact, this result is what one finds in actual
experiments.   

The experimental verdict on Gibbs' paradox has profound implications.   If we
could keep track of every molecule we would certainly have the extra entropy,
the so-called entropy of mixing.  Indeed, when gases of {\it different\/} types are
mixed, say hydrogen and helium, entropy {\it is\/} generated.  Since entropy of
mixing is not observed for (superficially) similar gases, there can be no method,
{\it even in principle}, to tell their molecules apart.  Thus we cannot make a
rigorous statement of the kind, ``Molecule 1 is in A, molecule 2 is in A,\lips ,
molecule
$n$ is in A'',  but only a much weaker statement, of the kind,  ``There are
$n$ molecules in A".      In this precise sense, hydrogen molecules are not only
similar, nor even only identical, but beyond that, indistinguishable.

Fermi motivated his concept of state-counting~\cite{statistics} through a
different, though related, difficulty of classical statistical mechanics, namely its
failure to account for Nernst's law.  This law, abstracted from empirical data,
implied that the entropy of a material vanishes at the absolute zero of
temperature.  Like Gibbs's paradox it indicates that there are far fewer states than
appear classically, and in particular no entropy of mixing.   Fermi therefore
proposed to generalize Pauli's exclusion principle from its spectroscopic origins
to a universal principle, not only for electrons but as a candidate to describe
matter in general:
\begin{quote} 
We will therefore assume in the following that, at most, one
molecule with given quantum numbers can exist in our gas: as quantum numbers
we must take into account not only those that determine the internal motions of
the molecule but also the numbers that determine its translational motion.
\end{quote} 
It is remarkable, and perhaps characteristic, that Fermi uses the
tested methods of the old quantum theory in evaluating the properties of his
ideal gas.  He places his molecules in an imaginary shallow harmonic well,
identifies the single-particle energy levels the levels by appeal to the
Bohr-Sommerfeld quantization rules, and assigns one state of the system to every
distribution of particles over these levels.   (Of course, he does not fail to remark
that the results should not depend on the details of this procedure.)  All the
standard results on the ideal Fermi-Dirac gas are then derived in a few strokes, by
clever combinatorics.  

Indeed, after a few introductory paragraphs, the paper becomes a series of
equations interrupted only by minimal text of the general form, ``now let us
calculate\plips. '' So the following interpolation, though brief, commands
attention:
\begin{quote} 
At the absolute zero point, our gas molecules arrange themselves
in a kind of shell-like structure which has a certain analogy to the arrangement of
electrons in an atom with many electrons.
\end{quote} 
We can  see in this the germ of the
Thomas-Fermi model of atoms~\cite{thomas-fermi}, which grows very directly
out of Fermi's treatment of the quantum gas, and gives wonderful insight into the
properties of matter~\cite{lieb}.   Also, of course, the general scheme of fermions
in a harmonic well is the starting point for the {\it nuclear\/} shell model -- of
which more below.  

The successful adaptation of Fermi's ideal gas theory to describe electrons in
metals, starting with Sommerfeld and Bethe,  and many other applications,
vindicated his approach.

\section{The Primacy of Quantum Fields 1: Free Fields}

As I have emphasized, Fermi's state-counting  logically requires the radical
indistinguishability of the underlying particles.  It does not explain that fact,
however, so its success only sharpens the fundamental problem posed in our
Question 1.   Deeper insight into this question requires considerations of another
order.  It comes from the synthesis of quantum mechanics and special relativity
into quantum field theory.

The field concept came to dominate physics starting with the work of Faraday in
the mid-nineteenth century.    Its conceptual advantage over the earlier
Newtonian program of physics, to formulate the fundamental laws in terms of
forces among atomic particles, emerges when we take into account the
circumstance,  unknown to Newton (or, for that matter, Faraday) but
fundamental in special relativity, that physical influences travel no faster than a
finite limiting speed.  For this implies that the force on a given particle at a given
time cannot be inferred from  the positions of other particles at that time, but
must be deduced in a complicated way from their previous positions.   Faraday's
intuition that the fundamental laws of electromagnetism could be expressed
most simply in terms of fields filling space and time was of course brilliantly
vindicated by Maxwell's mathematical theory. 

The concept of locality, in the crude form that one can predict the behavior of
nearby objects without reference to distant ones,  is basic to scientific practice.   
Practical experimenters -- if not astrologers -- confidently expect, on the basis of
much successful experience,  that after reasonable (generally quite modest)
precautions isolating their experiments, they will obtain reproducible results.

The deep and ancient historic roots of the field and locality concepts provide no
guarantee that these concepts remain relevant or valid when extrapolated far
beyond their origins in experience,  into the subatomic and quantum domain.   
This extrapolation must be judged by its fruits.    Remarkably, the first
consequences of relativistic quantum field theory supply the answer to our
Question 1, in its sharp form including Fermi's quantum state-counting.  

In quantum field theory, particles are not the primary reality.      Rather, as
demanded by relativity and locality, it is fields that are the primary reality. 
According to quantum theory, the
excitations of these fields come in discrete lumps.    These lumps are what we recognize as particles.   In this way, particles
are derived from fields.  Indeed, what we call particles are simply the form in
which low-energy excitations of quantum fields appear. Thus all electrons are
precisely alike because all are excitations of the same underlying Ur-stuff, the
electron field.  The same logic, of course, applies to photons or quarks, or even to
composite objects such as atomic nuclei, atoms, or molecules.  

Given the indistinguishability of a class of elementary particles, including
complete invariance of their interactions under interchange, the general
principles of quantum mechanics teach us that solutions forming any
representation of the permutation symmetry group retain that property in
time.   But they do not constrain which representations are realized.  Quantum
field theory not only explains the existence of indistinguishable particles and the
invariance of their interactions under interchange, but also constrains the
symmetry of the solutions.  There are two possibilities, bosons or fermions.   For
bosons, only the identity representation  is physical (symmetric wave functions);
for fermions, only the one-dimensional odd representation  is physical
(antisymmetric wave functions).    One also has the spin-statistics theorem,
according to which objects with integer spin are bosons, whereas objects with
half odd-integer spin are fermions.   Fermions, of course, obey Fermi's
state-counting procedure.   Examples are electrons, protons, neutrons, quarks,
and other charged leptons and neutrinos.  

It would not be appropriate to review here the rudiments of quantum field
theory, which justify the assertions of the preceding paragraph.  But a brief
heuristic discussion seems in order. 

In classical physics particles have definite trajectories, and there is no limit to
the accuracy with which we can follow their paths.  Thus, in principle, we could
always keep  tab on who's who.    Thus classical physics is inconsistent with the
rigorous concept of indistinguishable particles, and it comes out on the wrong
side of Gibbs' paradox.   

In the quantum theory of indistinguishable particles the situation is quite
different.   The possible positions of particles are  described by waves (i.e., their
wave-functions).  Waves can overlap and blur.   Related to this, there is a limit to
the precision with which their trajectories can be followed, according to
Heisenberg's uncertainty principle.      So when we calculate the
quantum-mechanical amplitude for a physical process to take place, we must
sum contributions from all ways in which it might have occurred.  Thus,
 to calculate the amplitude that a state with two indistinguishable
particles of a given sort  -- call them  quants -- at positions
$x_1, x_2$ at time
$t_i$, will evolve into a state with two quants at $x_3, x_4$ at time $t_f$, we must
sum contributions from all possible trajectories for the quants at intermediate
times.   These trajectories fall into two distinct categories.  In one category, the
quant initially at $x_1$ moves to $x_3$, and the quant initially at $x_2$ moves
to $x_4$.  In the other category, the quant initially at $x_1$ moves to $x_4$, and
the quant initially at
$x_2$ moves to $x_3$.  Because (by hypothesis) quants are indistinguishable,
the final states are the same for both categories. Therefore, according to the
general principles of quantum mechanics, we must add the amplitudes for these
two categories.    We say there are ``direct'' and  ``exchange'' contributions to
the process.    Similarly, if we have more than two quants, we must add
contributions involving arbitrary permutations of the original particles.    

Since the trajectories fall into discretely different classes, we may also consider
the possibility of adding the amplitudes with relative factors.  Mathematical
consistency severely limits our freedom, however.  We must demand that the
rule for multiplying amplitudes, when we sum over states at an intermediate
time, is consistent with the rule for the overall amplitude.  Since the final result
of double exchange is the same as no exchange at all, we must assign factors in
such a way that direct\ts$\times$\ts direct = exchange\ts$\times$\ts exchange, and the only
consistent possibilities are direct/exchange $= \pm 1$.    These correspond to
bosons ($+$) and fermions ($-$), respectively.   The choice of sign determines how
the interference term between direct and exchange contributions contributes to
the square of the amplitude, that is the probability for the overall process.   This
choice is vitally important for calculating the elementary interactions even of
short-lived, confined, or elusive particles, such as quarks and gluons, whose
equilibrium statistical mechanics is  moot~\cite{qualification}.

\section{The Centerpiece: Beta Decay}

The preceding consequences of quantum field theory follow from its basic
``kinematic'' structure, independent of any specific dynamical equations.   They
justified Fermi's state-counting, and rooted it in a much more comprehensive
framework.     But  Fermi's own major contributions to quantum field theory
came at the next stage, in understanding its dynamical implications.   

Fermi absorbed quantum electrodynamics by working through many examples
and using them in teaching.    In so doing, he assimilated Dirac's original, rather
abstract formulation of the theory to his own more concrete way of thinking.  
His review article~\cite{RMP} is a masterpiece, instructive and refreshing to read
even today.  It begins
\begin{quote} 
Dirac's theory of radiation is based on a very simple idea; instead
of considering an atom and the radiation field with which it interacts as two
distinct systems, he treats them as a single system whose energy is the sum of
three terms: one representing the energy of the atom, a second representing the
electromagnetic energy of the radiation field, and a small term representing the
coupling energy of the atom and the radiation field\plips. 
\end{quote} 
and soon continues
\begin{quote} 
A very simple example will explain these relations.  Let us consider
a pendulum which corresponds to the atom, and an oscillating string in the
neighborhood of the pendulum which represents the radiation field\plips. To
obtain a mechanical representation of this [interaction] term, let us tie the mass
$M$ of the pendulum to a point $A$ of the string by means of a very thin and
elastic thread $a$\plips\linebreak[3] [I]f a period of the of the string is equal to
the period of the pendulum, there is resonance and the amplitude of vibration of
the pendulum becomes considerable after a certain time.  This process
corresponds to the absorption of radiation by the atom.
\end{quote} 
Everything is done from scratch, starting with harmonic oscillators. 
Fully elaborated examples of how the formalism reproduces concrete
experimental arrangements in space-time are presented, including the Doppler
effect and Lippmann fringes, in addition to the  ``S-matrix'' type scattering
processes that dominate modern textbooks.   

Ironically, in view of what was about to happen, Fermi's review article does not
consider systematic quantization of the electron field.   Various processes
involving positrons are discussed on an {\it ad hoc\/} basis, essentially following
Dirac's hole theory.  

With hindsight, it appears obvious that Chadwick's discovery of the neutron in
early 1932 marks the transition between the ancient and the classic eras of
nuclear physics.    (The dominant earlier idea, consonant with an application of
Occam's razor that in retrospect seems reckless, was that nuclei were made from
protons and tightly bound electrons.   This idea is economical of particles, of
course, but it begs the question of dynamics, and it also has problems with
quantum statistics -- N$^{14}$ would be 21 particles, and a fermion, whereas
molecular spectroscopy shows it is a boson.)    At the time, however, there was
much confusion~\cite{pais}. 

The worst difficulties, which brought into question the validity of quantum
mechanics and energy conservation in the nuclear domain, concerned $\beta$
decay.  On the face of it, the observations seemed to indicate a process
$n\rightarrow p + e^-$, wherein a neutron decays into a proton plus electron. 
However, this would indicate boson statistics for the neutron, and it reintroduces
the spectroscopic problems mentioned above.   Moreover, electrons are 
observed to be emitted with a nontrivial spectrum of energies, which is
impossible for a two-body decay if energy and momentum are conserved.  Bohr
in this context suggested abandoning conservation of energy.   However, Pauli
suggested that the general principles of quantum theory could be respected and
the conservation laws could be maintained if the decay were in reality $n
\rightarrow p + e^- + {\bar \nu}$, with a neutral particle $\bar \nu$ escaping
detection.   

Fermi invented the term ``neutrino'' for Pauli's particle.  This
started as a little joke in conversation,  \emph{neutrino} being the Italian
diminutive of \emph{neutrone},  suggesting the contrast of a little neutral one
versus a heavy neutral one.  It stuck, of course.  (Actually, what appears in
neutron decay is what we call today an antineutrino.) 
  
More importantly, Fermi took Pauli's idea seriously and literally, and attempted
to bring it fully into line with special relativity and quantum mechanics.  This
meant constructing an appropriate quantum field theory.  Having mastered the
concepts and technique of quantum electrodynamics, and after absorbing the
Jordan-Wigner technique for quantizing fermion fields,  he was ready to
construct a quantum field theory for $\beta$ decay.  He chose a Hamiltonian of
the simplest possible type, involving a local coupling of the  four fields involved,
one for each particle created or destroyed.   There are various possibilities for
combining the  spinors  into a  Lorentz-invariant object, as Fermi discusses.   He
then calculates the electron emission spectrum, including the possible effect of
nonzero neutrino mass.   

It soon became apparent that these ideas successfully organize a vast wealth of
data on nuclear
$\beta$ decays in general.  They provided, for forty years thereafter, the
ultimate foundation of weak interaction theory, and still remain the most useful
working description of a big chapter of nuclear and particle physics.   Major
refinements including the concept of universality, parity violation, V-A theory --
which together made the transition to a deeper foundation, based on the gauge
principle  compelling -- not to mention the experimental investigation of
neutrinos themselves, were all  based firmly on the foundation supplied by
Fermi.

\section{The Primacy of Quantum Fields 2:\protect\\ From Local
Interactions to Particles/Forces\protect\\ (Real and Virtual)}

Although Fermi's work on $\beta$ decay was typically specific and sharply
focused, by implication it set a much broader agenda for quantum field theory. 
It emphasized the very direct connection between the abstract principles of
interacting quantum field theory and a most fundamental aspect of Nature,  the
{\it the ubiquity of particle creation and destruction processes}.

Local interactions involve products of field operators at a point.   When the fields
are expanded into creation and annihilation operators multiplying modes,  we
see that these interactions correspond to processes wherein particles can be
created, annihilated, or changed into different kinds of particles.   This
possibility arises, of course, in the primeval quantum field theory, quantum
electrodynamics, where the primary interaction arises from a product of the
electron field, its Hermitean conjugate,  and the photon field.  Processes of
radiation and absorption of photons by electrons (or positrons), as well as
electron-positron pair creation,  are encoded in this product.   But because the
emission and absorption of light is such a common experience, and
electrodynamics such a special and familiar classical field theory,  this
correspondence between formalism and reality did not initially make a big
impression.  The first conscious exploitation of the potential for quantum field
theory to  describe processes of transformation was Fermi's theory of beta
decay.   He turned the procedure around,  inferring from the observed processes
of particle transformation the nature of the underlying local interaction of
fields.    Fermi's theory involved creation and annihilation not of photons but of
atomic nuclei and electrons (as well as neutrinos)  -- the ingredients of
``matter''.     It thereby initiated the process whereby classical atomism,
involving stable individual objects,  was replaced by a more sophisticated and
accurate picture.  In this picture only the fields,  and not the individual objects
they create and destroy,  are permanent.

This line of thought gains power from its association with a second general
consequence of quantum field theory,  {\it the association of forces and
interactions with particle exchange}.    When Maxwell completed the equations of
electrodynamics, he found that  they supported source-free electromagnetic
waves.   The classical electric and magnetic fields took on a life of their own.  
Electric and magnetic forces between charged particles are explained as due  to
one particle acting as a source for electric and magnetic fields, which then
influence others.     With the correspondence of fields and particles, as it arises
in quantum field theory, Maxwell's discovery corresponds to the existence of
photons,  and the generation of forces by intermediary fields corresponds to the
exchange of virtual photons.   

The association of forces (or more generally, interactions) with particles is a
general feature of quantum field theory.  ``Real'' particles are field excitations
that can be  considered usefully as independent entities, typically because they
have a reasonably long lifetime and can exist spatially separated from other
excitations, so that we can associate transport of definite units of mass, energy,
charge, etc., with them.  But in quantum theory all excitations can also be
produced as short-lived fluctuations.  These fluctuations are constantly taking
place in what we regard as empty space, so that the physicists' notion of vacuum
is very far removed from simple nothingness.   The behavior of real particles is
affected by their interaction with these virtual fluctuations.  Indeed, according to
quantum field theory that's all there is!   So observed forces must be ascribed to
fluctuations of quantum fields -- but these fields will then also support genuine
excitations, that is, real particles.   Tangible particles and their ``virtual''
cousins  are as inseparable as two sides of the same coin.  This connection was
used by Yukawa to infer the existence and mass of pions from the range of
nuclear forces.  (Yukawa began his work by considering whether the exchange of
virtual electrons and neutrinos, in Fermi's $\beta$ decay theory, might be
responsible for the nuclear force!  After showing that these virtual particles gave
much too small a force, he was led to a new particle.)  More recently it has been
used in electroweak theory to infer the existence, mass,  and properties of $W$
and $Z$ bosons prior to their observation, and in QCD to infer the existence and
properties of gluon jets prior to their observation.

This circle of ideas, which to me forms the crowning glory of twentieth-century
physics, grew around Fermi's theory of $\beta$ decay. There is a double
meaning in my title, ``Fermi and the Elucidation of Matter''.  For Fermi's most
beautiful insight was, precisely, to realize the profound commonality of matter
and light.

\section{Nuclear Chemistry}

The other big classes of nuclear transformations, of quite a different character
from $\beta$ decay, are those in which no leptons are involved.  In modern
language, these are processes mediated by the strong and electromagnetic
interactions.  They include the fragmentation of heavy nuclei (fission) and the
joining of light nuclei (fusion).    These processes are sometimes called nuclear
chemistry, since they can be pictured as rearrangements of existing materials --
protons and neutrons -- similar to how ordinary chemistry can be pictured as
rearrangements of electrons and nuclei.   In this terminology, it would be natural
to call $\beta$ decay nuclear alchemy.  

Fermi discovered the technique that above all others opened up the experimental
investigation of nuclear chemistry.  This is the potent ability of slow neutrons to
enter and stir up nuclear targets.  Fermi regarded this as his greatest discovery. 
In an interview with S.~Chandrasekhar, quoted in~\cite{segre}, he described it:
\begin{quote} 
I will tell you now how I came to make the discovery which I
suppose is the most important one I have made.  We were working very hard on
the neutron-induced radioactivity and the results we were obtaining made no
sense.  One day, as I came to the laboratory, it occurred to me that I should
examine the effect of placing a piece of lead before the incident neutrons. 
Instead of my usual custom, I took great pains to have the piece of lead precisely
machined.  I was clearly dissatisfied with something: I tried every excuse to
postpone putting the piece of lead in its place.  When finally, with some
reluctance, I was going to put it in its place, I said to myself: ``No, I do not want
this piece of lead here; what I want is a piece of paraffin.'' It was just like that,
with no advance warning, no conscious prior reasoning.  I immediately took some
odd piece of paraffin and placed it where the piece of lead was to have been.
\end{quote} 
There are wonderful accounts of how he mobilized his group in
Rome to  exploit, with extraordinary joy and energy, his serendipitous
discovery~\cite{segre}. 

I will not retell that story here, nor the epic saga of the nuclear technology
commencing with the atomic bomb project~\cite{Rhodes}.  We are still coming to
terms with the destructive potential of nuclear weapons, and we have barely
begun to exploit the resource of nuclear energy. 

From the point of view of pure physics, the significance of Fermi's work in
nuclear chemistry was above all to show that in a  wealth of phenomena nuclei
could be usefully described as composites wherein protons and neutrons, though
in close contact, retain their individual identity and properties.   This viewpoint
reached its apex in the shell model of Mayer and Jensen~\cite{shell}.   Famously, 
Fermi helped inspire Mayer's work, and in particular suggested the importance of
spin-orbit coupling, which proved crucial.    The success of Mayer's
independent-particle model for describing quantum systems in which the
interactions are very strong stimulated deep work in the many-body problem. It
also provided the intellectual background for the quark model.  
\bigskip

\section{Last Insights and Visions}

With developments in nuclear chemistry making it clear that different nuclei are
built up from protons and neutrons,  and developments in beta-decay theory
showing how protons and neutrons can transmute into one another, our
Question 2, to understand the constitution of the world, came into sharp focus. 
Specifically, it became possible to pose the origin of the elements as a scientific
question.  Fermi was very much alive to this possibility that his work had opened
up.   With Turkevich he worked extensively on Gamow's ``ylem'' proposal that
the elements are built up by successive neutron captures, starting with neutrons
only in a hot, rapidly expanding Universe.   They correctly identified the major
difficulty with this idea, that is, the insurmountable gap at atomic number 5,
where no suitably stable nucleus exists.   Yet a rich and detailed account of the
origin of elements can be constructed, very nearly along these lines.   The
Fermi-Turkevich gap at $A=5$ is reflected in the observed abundances,  in that
less than 1\% of  cosmic chemistry resides in such nuclei, the astronomers'
``metals''.  Elements beyond
$A=5$ (except for a tiny amount of
Li$^7$) are produced in a different way, in the reactions powering stars, and are
injected back into circulation through winds or, perhaps, with additional
last-minute cooking, during supernova explosions.  Also, the correct initial
condition for big-bang nucleosynthesis postulates thermal equilibrium at high
temperatures, not all neutrons.   Though he didn't quite get there,  Fermi
envisioned this promised land.

All this progress in observing, codifying, and even controlling nuclear processes
was primarily based on experimental work.  The models used to correlate the
experimental data incorporated relativistic kinematics and basic principles of
quantum mechanics, but were not derived from a closed set of underlying
equations.   The experimental studies reveal that the interaction between
nucleons at low energy is extremely complex.  It depends on distance, velocity,
and spin in a completely entangled way.   One could parametrize the
experimental results in terms of a set of energy-dependent functions -- ``phase
shifts'' -- but  these functions displayed no obvious simplicity.     

The lone triumph of high-energy theory was the discovery of Yukawa's pion. 
This particle, with the simple local coupling postulated by  Yukawa's theory,
could account semiquantitatively for the long-range tail of the force.   Might it
provide the complete answer?  No one could tell for sure -- the necessary
calculations were too difficult. 

Starting around 1950 Fermi's main focus was on the experimental investigation
of pion-nucleon interactions.  They might be expected to have only the square root
of the difficulty in interpretation, so to speak, since they are closer to the
core-element of Yukawa's theory.  But pion-nucleon interactions turned out to
be  extremely complicated also.  

With the growing complexity of the observed phenomena, Fermi began to doubt
the adequacy of Yukawa's theory.  No one could calculate the consequences of
the theory accurately, but the richness of the observed phenomena undermined
the basis for hypothesizing that with point-like protons, neutrons, and pions one
had reached bottom in the understanding of strongly interacting matter.   There
were deeper reasons for doubt, arising from a decent respect for Question 2. 
Discovery of the $\mu$ particle, hints of more new particles in cosmic ray events
(eventually evolving into our $K$ mesons),  together with the familiar nucleons,
electrons, photons, plus neutrinos and the pions, indicated a proliferation of
``elementary'' particles.  They all transformed among one another in
complicated ways.  Could one exploit the transformative aspect of quantum field
theory to isolate a simple basis for this proliferation -- fewer, more truly
elementary building blocks?  

In one of his late theoretical works, with Yang, Fermi proposed a radical
alternative  to Yukawa's theory, that might begin to economize particles.   They
proposed that the pion was not fundamental and elementary at all, but rather a
composite particle, specifically a nucleon-antinucleon bound state.   This was a
big extrapolation of the idea behind the shell model of nuclei.   Further, they
proposed that the primary strong interaction was what we would call today a
four-fermion coupling of the nucleon fields.  The pion was to be produced as a
consequence of this interaction, and Yukawa's theory as an approximation --
what we would call today an effective field theory.   The primary interaction in
the Fermi-Yang theory is of the same form of interaction as appears in Fermi's
theory of $\beta$ decay, though of course the strength of the interaction and the
identities of the fields involved are quite different.   In his account of this
work~\cite{yang} Yang says
\begin{quote}
 As explicitly stated in the paper, we did not really have any
illusions that what we suggested may actually correspond to reality\plips. Fermi
said, however, that as a student one solves problems, but as a research worker
one raises questions\plips. 
\end{quote} 
Indeed, the details of their proposal do not correspond to our
modern understanding.  In particular, we have learned to be comfortable with a
proliferation of particles, so long as their fields are related by symmetries.  But
some of the questions Fermi and Yang raised -- or,  I would say, the directions
they implicitly suggested -- were, in retrospect, fruitful ones.   First, the whole
paper is firmly set in the framework of relativistic quantum field theory.   Its
goal, in the spirit of quantum electrodynamics and Fermi's $\beta$ decay
theory,  was to explore the possibilities of that framework, rather than to
overthrow it.  For example, at the time the existence of antinucleons had not yet
been established experimentally.  However, the existence of antiparticles is a
general consequence of relativistic quantum field theory, and it is accepted with
minimal comment.  Second, building up light particles from much heavier
constituents was a liberating notion.  It is an extreme extrapolation of the
lowering of mass through binding energy.   Nowadays we go much further along
the same line, binding infinitely heavy (confined) quarks and gluons into the
observed strongly interacting particles, including both pions and nucleons. 
Third, and most profoundly, the possibility of deep similarity in the basic
mechanism of the strong and the weak interaction, despite their very different
superficial appearance, is  anticipated.    The existence of just such a common
mechanism, rooted in concepts later discovered by Fermi's co-author --
Yang-Mills theory -- is a central feature of the modern theory of matter, the
so-called Standard Model.

In another of his last works, with Pasta and Ulam~\cite{FPU}, Fermi
enthusiastically seized upon a new opportunity for exploration -- the emerging
capabilities of rapid machine computation.   With his  instinct for the border of
the unknown and the accessible, he chose to revisit a classic, fundamental
problem that had been the subject of one of his earliest papers, the problem of
approach to equilibrium in a many-body system.   The normal working
hypothesis of statistical mechanics is that equilibrium is the default option
and is rapidly attained in any complex system unless some simple conservation
law forbids it.  But proof is notoriously elusive, and Fermi   thought to explore the
situation by controlled {\it numerical \/} experiments, wherein the degree of 
complexity could be varied.  Specifically, he coupled together various modest
numbers of locally coupled nonlinear springs.   A stunning surprise emerged:
The approach to equilibrium is far from trivial; there are emergent structures,
collective excitations that can persist indefinitely.    The topic of solitons, which
subsequently proved vast and fruitful, was foreshadowed in this work.  And the
profound but somewhat nebulous question, central to an adequate
understanding of Nature, of how ordered structures emerge spontaneously from
simple homogeneous laws and minimally structured initial conditions,
spontaneously began to pose itself.   

Not coincidentally, the same mathematical structure -- locally coupled
nonlinear  oscillators -- lies at the foundation of relativistic quantum field
theory.   Indeed, as we saw, this was the way Fermi approached the subject  right
from the start.   In modern QCD  the emergent structures are protons, pions, and
other hadrons, which are well hidden in the quark and gluon field ``springs''.   
Numerical work of the kind Fermi pioneered remains our most reliable tool for
studying these structures.

Clearly, Fermi was leading physics toward fertile new directions.  When his life
was cut short, it was a tremendous loss for our subject.

\section{Fermi as Inspiration: Passion and Style}

Surveying Fermi's output as a whole, one senses a special passion and style,
unique in modern physics.      Clearly, Fermi loved his dialogue with Nature.    He
might respond to her deepest puzzles with queries of his own, as in the
explorations we have just discussed, or with patient gathering of facts, as in his
almost brutally systematic experimental investigations of nuclear transmutations
and pion physics.  But, he also took great joy in solving, or simply explicating, her
simpler riddles, as the many Fermi stories collected in this volume attest.   

Fermi tackled ambitious problems at the frontier of knowledge, but always with
realism and modesty. These aspects of his scientific style shine through one of
Fermi's rare ``methodological'' reflections, written near the end of his life~\cite{method}:

\begin{quote}
When the Yukawa theory first was proposed there was a legitimate hope that the
particles involved, protons, neutrons, and pi-mesons,  could be legitimately
considered as elementary particles.  This hope loses more and more its
foundation as new elementary particles are rapidly being discovered.

It is difficult to say what will be the future path.  One can go back to the books on
method  (I doubt whether many physicists actually do this) where it will be
learned that one must take experimental data, collect experimental data,
organize experimental data, begin to make working hypotheses, try to correlate,
and so on, until eventually a pattern springs to life and one has only to pick out
the results.  Perhaps the traditional scientific method of the text books may be
the best guide, in the lack of anything better\plips.  

Of course, it may be that someone will come up soon with a solution to the
problem of the meson,  and that experimental results will confirm so many
detailed features of the theory that it will be clear to everybody that it is the
correct one. Such things have happened in the past.  They may happen again. 
However, I do not believe that we can count on it, and I believe that we must be
prepared for a long, hard pull.
\end{quote}

Those of you familiar with the subsequent history of the strong-interaction
problem will recognize that Fermi's prognosis was uncannily  accurate.  A long
period of experimental exploration and pattern-recognition provided the
preconditions for a great intellectual leap and synthesis.   The process would, I
think, have gratified Fermi, but not surprised him.  

The current situation in physics is quite different from what Fermi lived through
or, at the end, described.  After the triumphs of the twentieth century, it is easy
to be ambitious.  Ultimate questions about the closure of fundamental dynamical
laws and the origin of the observed Universe begin to seem accessible.    The
potential of quantum engineering, and the challenge of understanding how one
might orchestrate known fundamentals into complex systems, including powerful
minds, beckon.   Less easy, perhaps, is to remain realistic and (therefore)
appropriately modest -- in other words, to craft important subquestions that we
can answer definitively, in dialogue with Nature.  In this art Fermi was a natural
grand master, and the worthy heir of Galileo.

\subsubsection*{Acknowledgments} This work is supported in part by funds
provided by the U.S. Department of Energy (D.O.E.) under cooperative research
agreement
\#DF-FC02-94ER40818.  

I would like to thank  Marty Stock for help with the
manuscript.


\begin{thebibliography}{99}


\bibitem{works}{\it Collected Works}, 2 volumes, E. Fermi (University of Chicago
Press, 1962).

\bibitem{segre}Another indispensable source is the scientific biography {\it
Enrico Fermi, Physicist}, E. Segre (University of Chicago Press, 1970). 
 
\bibitem{atoms}An accessible selection of original sources and useful
commentaries is {\it The World of the Atom}, 2 volumes, ed. H. Boorse and L.
Motz (Basic Books, 1966).

\bibitem{statistics}Papers 30 and 31 in~\cite{works}.  A portion is translated into
English in~\cite{atoms}. 

\bibitem{thomas-fermi}Papers 43-48 in~\cite{works}.  

\bibitem{lieb}{\it Thomas-Fermi and Related Theories of Atoms and Molecules},
E. Lieb, in Rev. Mod. Phys. {\bf 53} No. 4, pt. 1, 603-641 (1981).

\bibitem{qualification}Actually, the statistical mechanics of quarks and gluons
may become accessible in ongoing experiments at RHIC.  

\bibitem{RMP}Paper 67 in~\cite{works}.

\bibitem{pais}{\it Inward Bound: Of Matter and Forces in the Physical World}, A.
Pais (Oxford, 1986).

\bibitem{Rhodes}{\it The Making of the Atomic Bomb}, R. Rhodes (Simon and
Schuster, 1986). 

\bibitem{shell}{\it Elementary Theory of Nuclear Shell Structure}, M. Mayer
(Wiley, 1955).

\bibitem{yang}C. N. Yang, introduction to paper 239 in~\cite{works}. 

\bibitem{FPU}Paper 266 in~\cite{works}. 

\bibitem{method}Paper 247 in~\cite{works}. 

\end{thebibliography}
\end{document}